\documentclass{article}
\usepackage[top=30truemm,bottom=30truemm,left=25truemm,right=25truemm]{geometry}
\usepackage{amsmath,amssymb}
\usepackage{bm}
\usepackage{graphicx} 
\usepackage{wrapfig}

\def\NAME#1#2{\caption{#2} \label{#1}}

\def\SEC#1#2{\section{#1} \label{#2}}

\def\FIGT#1{\begin{figure}[!t] \begin{center} #1 \end{center} \end{figure}}

\def\TABLEB#1{\begin{table}[!b] \begin{center} #1 \end{center} \end{table}}

\def\EPSH#1#2#3{\includegraphics[height= #2 cm,clip]{#1.eps} \NAME{#1}{#3}}
\def\EPSW#1#2#3{\includegraphics[width= #2 cm,clip]{#1.eps} \NAME{#1}{#3}}

\def\EPS08#1#2#3#4{\includegraphics[#3= #4 cm,clip]{../eps08/#1.eps} \NAME{#1}{#2}}

\def\EQ#1{\begin{equation} #1 \end{equation}}

\def\EQL#1#2{\begin{equation} \label{#1} #2 \end{equation}}
\def\EQN#1{\begin{eqnarray} #1 \end{eqnarray}}

\def\SL{\left(}
\def\SR{\right)}

\def\LRA#1{ \langle #1 \rangle}

\newfont{\bg}{cmr10 scaled\magstep5}
\newfont{\bbg}{cmsy10 scaled\magstep5}

\newcommand{\bigzerou}{\smash{\lower2.7ex\hbox{\bg 0}}}

\newcommand{\bbR}{{\mathbb R}}

\usepackage{multirow}

\title{Coordination game in bidirectional flow}

\author{
Daichi Yanagisawa \\
Research Center for Advanced Science and Technology, The University of Tokyo,\\4-6-1, Komaba, Meguro-ku, Tokyo, 153-8904, Japan \\
tDaichi@mail.ecc.u-tokyo.ac.jp
}

\date{}

\begin{document}

\maketitle

\begin{abstract}
We have introduced evolutionary game dynamics to a one-dimensional cellular-automaton to investigate evolution and maintenance of cooperative avoiding behavior of self-driven particles in bidirectional flow.
In our model, there are two kinds of particles, which are right-going particles and left-going particles. 
They often face opponent particles, so that they swerve to the right or left stochastically in order to avoid conflicts.
The particles reinforce their preferences of the swerving direction after their successful avoidance.
The preference is also weakened by memory-loss effect.

Result of our simulation indicates that cooperative avoiding behavior is achieved, i.e., swerving directions of the particles are unified, when the density of particles is close to 1/2 and the memory-loss rate is small.
Furthermore, when the right-going particles occupy the majority of the system, we observe that their flow increases when the number of left-going particles, which prevent the smooth movement of right-going particles, becomes large.
It is also investigated that the critical memory-loss rate of the cooperative avoiding behavior strongly depends on the size of the system.
Small system can prolong the cooperative avoiding behavior in wider range of memory-loss rate than large system.

\end{abstract}

\SEC{Introduction}{INTRO}

Coordination game is a class of games in game theory where Nash equilibria are achieved when the players choose the same strategy.
It is applied to study the choice of technological standards \cite{Mattli2003wp}, tax compliance \cite{Bloomquist2011pfr}, and trading behavior \cite{Bian2016pa}.

\FIGT{\EPSW{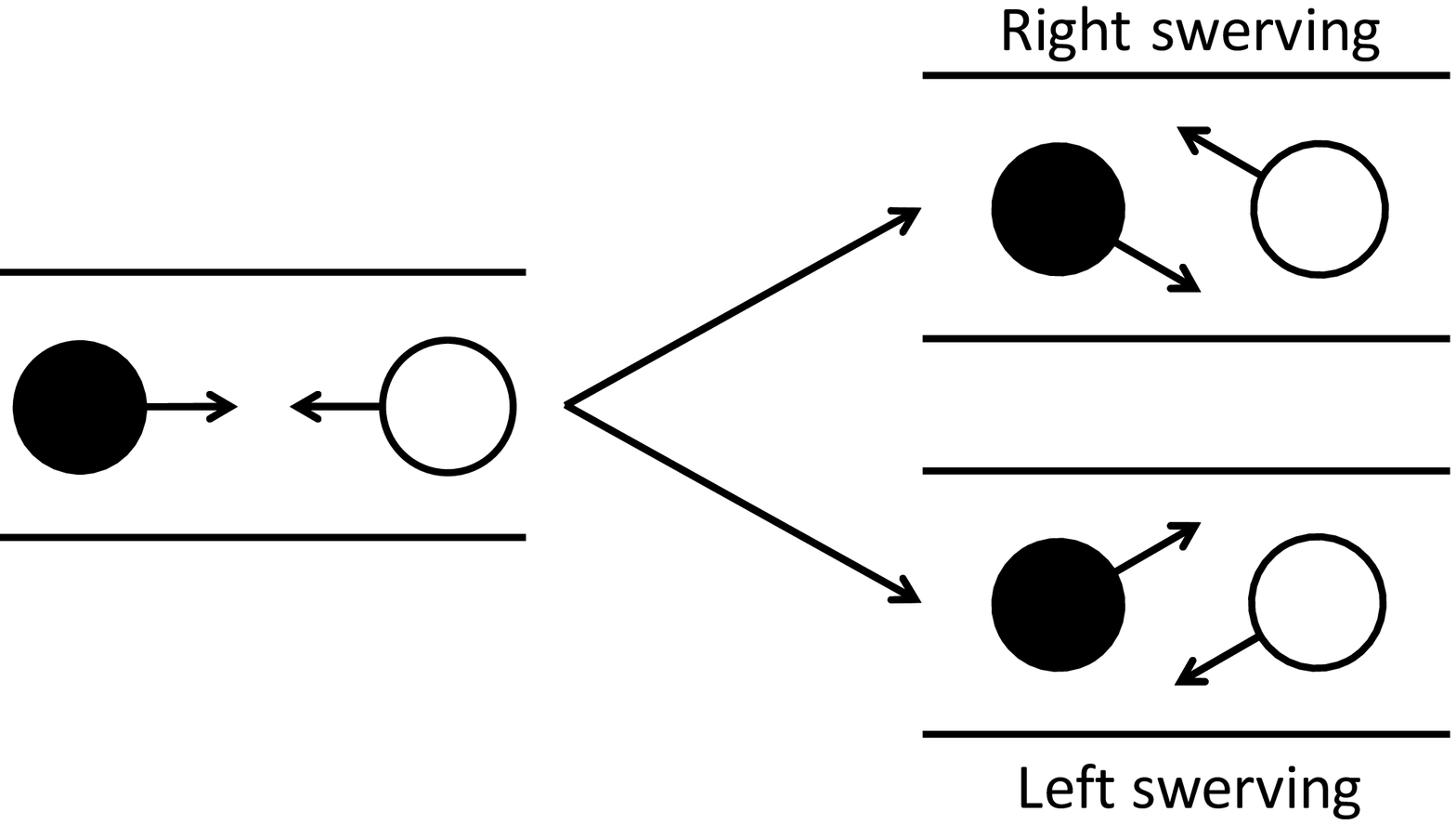}{10}{
Schematic view of collision avoidance by swerving in bidirectional flow.
}}

We see coordination-game-like phenomena in pedestrian dynamics, which has been vigorously studied in these two decades \cite{Helbing2001, Seyfried2005, Antonini2006, Yanagisawa2009, Schadschneider2010}.
Collision avoidance by swerving in bidirectional flow, which has been investigated both theoretically \cite{Helbing1995, Blue2001, Flotterod2015, Morton2016} and experimentally \cite{Hoogendoorn2005, Zhang2013a, Feliciani2015a} is a representative example.
When a right-going pedestrian face to a left-going pedestrian as in Fig. \ref{f12_swerving_direction}, he/she has to avoid the opponent by swerving to the right or left.
If the swerving directions of the two pedestrians do not agree, they need to adjust their strategy (right or left) by trial and error.
Thus, the unification of the swerving directions of pedestrians smooths bidirectional flow and has a positive effect as the choice of technological standards. 

This correspondence between coordination game and collision avoidance reminds us that we need to consider frequency of interaction among pedestrians when we study the cooperative avoiding behavior in pedestrian dynamics.
Walking side of pedestrians is not strictly determined as that of vehicles, and it is presumable to consider that pedestrians learn appropriate swerving direction in their living culture and society.
Thus, frequency of interaction (chance for learning swerving direction) is an important factor for the unification of the swerving directions.

Besides, original game theory does not include spatial effect.
Recently, evolutionary coordination game has been studied on networks \cite{Bian2016pa, Konno2015}; however, players in such models stay at the nodes and do not move in the space, in other words the nodes represent the players.
In the real world, players, such as pedestrians, move by their selves and interact with each other.

With the two motivations above, we develop a new model by combining a one-dimensional cellular automaton and evolutionary game dynamics in order to investigate evolutionary game dynamics with moving particles on a lattice.
Particles in the model have memories of their preferred swerving directions, which are updated by interaction with other particles and memory-loss effect.
Jam in the lattice deprives the particles of interactions with others; therefore, the spatial effect on the frequency of interaction is also studied. 
Although we obtain the motivation of this research from bidirectional flow in pedestrian dynamics, elucidation of evolutionary coordination-game dynamics on a lattice is the main goal in this paper.

The remainder of this paper is organized as follows.
In the next section, our model is introduced in detail.
In Sec. \ref{EQUAL}, we study how the density and memory-loss rate affect the cooperative avoiding behavior when the number of right-going and left-going particles are same.
Subsequently, we consider the case where the number of right-going and left-going particles are different in Sec. \ref{DIFF}.
It is shown that increase of opponent particles, which seems to disrupt the smooth movement, improves the flow.
In Sec. \ref{SIZE}, we investigate how the size of the system affect the condition of the cooperative avoiding behavior by simulation and approximate analysis.
The final section is devoted to summary and conclusion.
\SEC{Model}{MODEL}
\FIGT{\EPSW{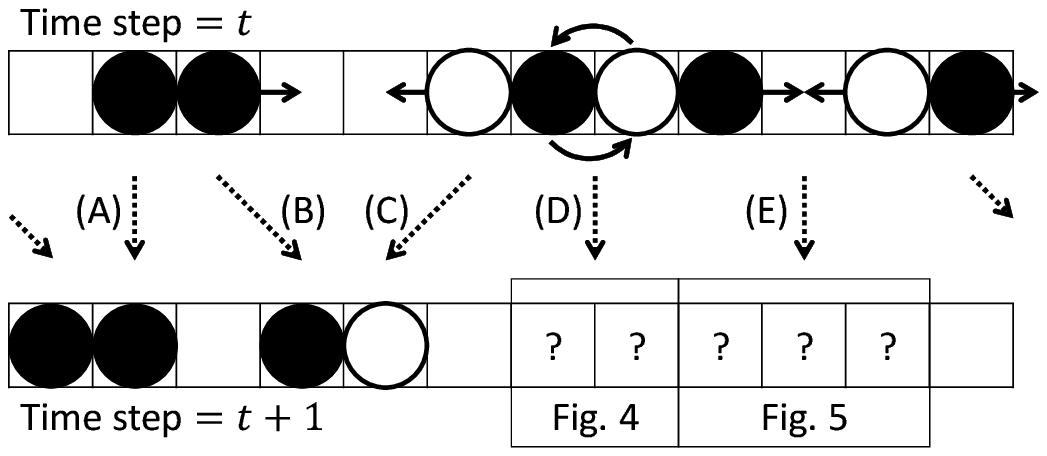}{12}{
Schematic view of the model.
(A) The black particle cannot move since its target cell is occupied by the other particle moving in the same direction.
(B) The black particle moves to the vacant right cell.
(C) The white particle moves to the vacant left cell.
(D) and (E) Interaction between the black and white particles occurs.
}}

A schematic view of our model is depicted in Fig. \ref{f3_model}.
We consider one-dimensional discrete space with periodic boundary condition.
The size of the system, i.e., the number of cell in the system, is $L$.
Time is also discrete in the model.

There are two kinds of particles, which are right-going (black) and left-going (white) particles.
The number of the right-going and left-going particles are $N_{\rm{R}}$ and $N_{\rm{L}}$, respectively.
Similarly, the density of the right-going and left-going particles are $\rho_{\rm{R}}=N_{\rm{R}}/L$ and $\rho_{\rm{L}}=N_{\rm{L}}/L$, respectively.
The total number of the particles is $N=N_{\rm{R}}+N_{\rm{L}}$.

Every discrete time step, first, all the right-going particles are updated in parallel, and then, all the left going particles are updated in parallel.
Each right-going (left-going) particle moves to the right (left) for one cell if their target cell is vacant (Fig. \ref{f3_model} case (B) and (C)).
It cannot move if their target cell is occupied by the particles moving in the same direction (Fig. \ref{f3_model} case (A)).

When the right-going and left-going particles exchange their position as in Fig. \ref{f3_model} case (D), they try to avoid each other by swerving to the right or left with the probabilities $p_i$ or $1-p_i$, respectively (Fig. \ref{f5_swerve2}), where $p_i$ is the right-swerving probability of the particle $i \in [1, N]$.
If the swerving directions of the two particles agree with the probability $p_i p_j + (1-p_i)(1-p_j)$ ($i, j \in [1, N]$, $i \not = j$), they avoid a conflict and exchange their position (Fig. \ref{f5_swerve2} case (A), (B)).
In contrast, when the swerving directions disagree with the probability $p_i(1-p_j) + p_j(1-p_i)$, a conflict occurs and they remain at their cell (Fig. \ref{f5_swerve2} case (C), (D)).

In the case where the right-going and left-going particles are trying move to the same cell as in Fig. \ref{f3_model} case (E), first, the right-going particle moves one cell due to the updating order.
Then, the left-going particle tries to penetrate into the cell occupied by the right-going particle.
The similar rule used in case (D) is exploited to judge avoidance and conflict.
The details are summarized in Fig. \ref{f4_swerve1}.
Note that the advantage of the updating order for the right-going particles does contribute to the differences between the results of right-going and left-going particles when the simulated time steps is short; however, if we perform a simulation long enough and consider average values, the differences are neglected.

\FIGT{\EPSW{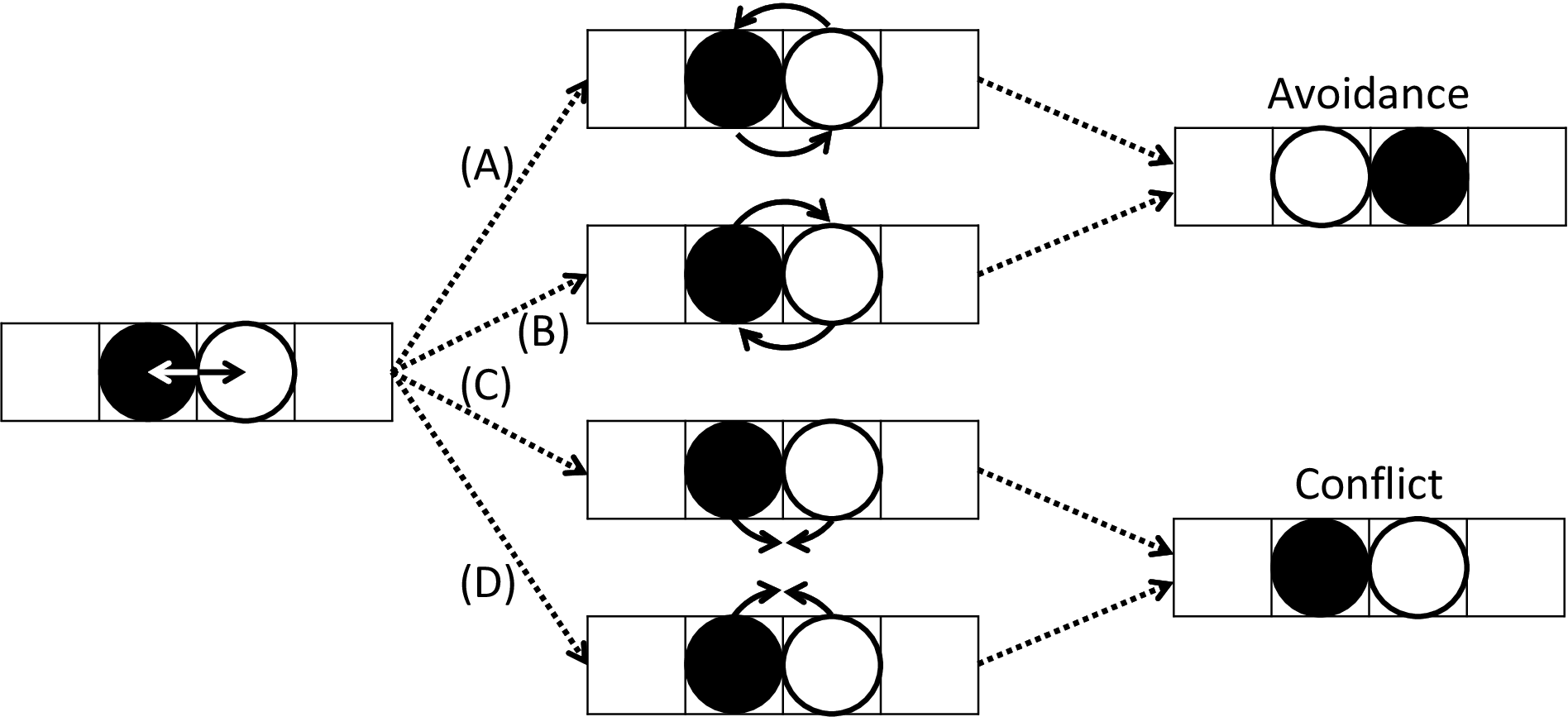}{9.8}{
Schematic view of avoidance and conflict when the two particles try to exchange their cell.
(A) Avoidance achieved by right swerving.
(B) Avoidance achieved by left swerving.
(C) and (D) Conflict.
}}

\FIGT{\EPSW{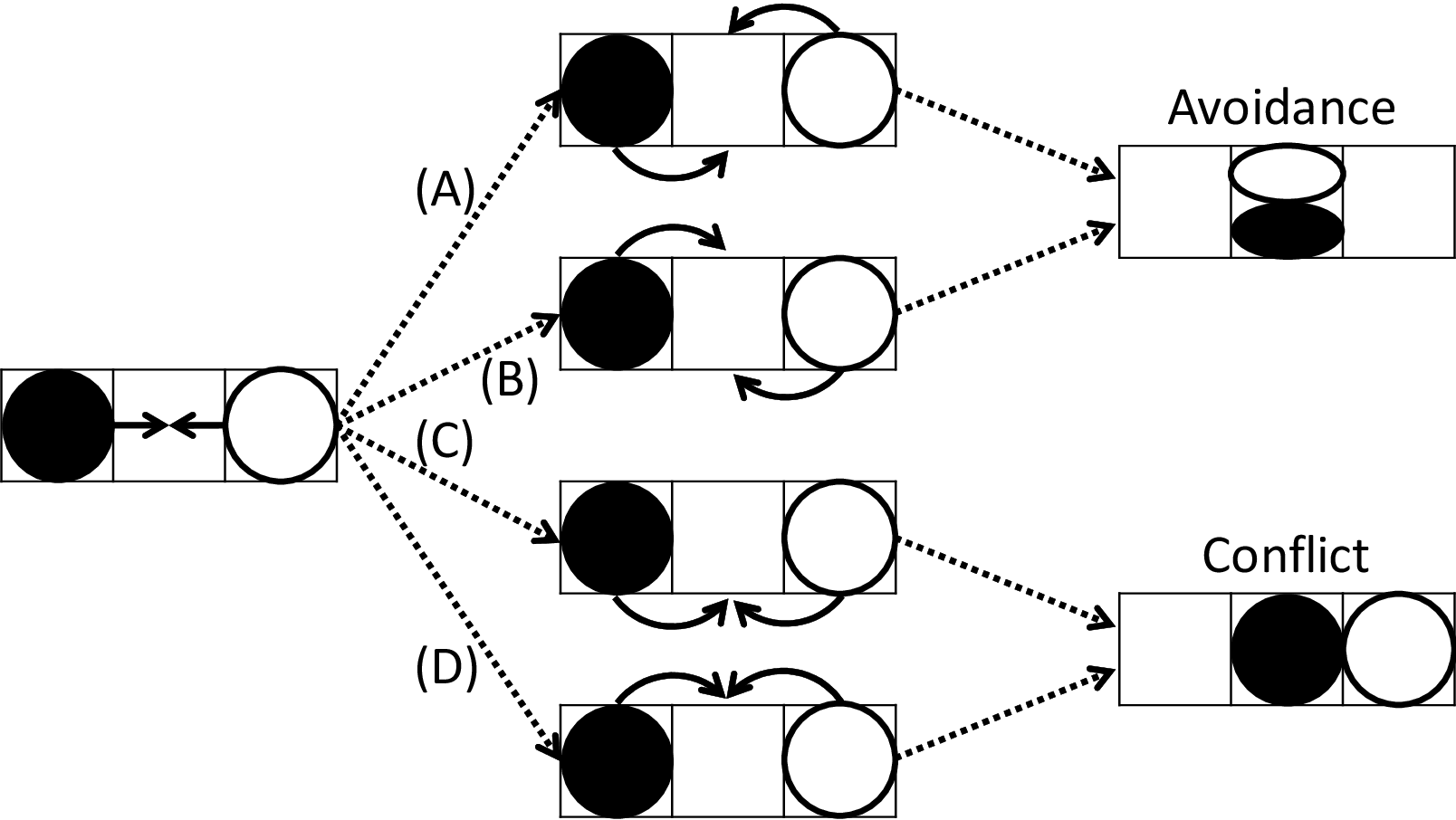}{8}{
Schematic view of avoidance and conflict when the two particles move to the same cell.
(A) Avoidance achieved by right swerving.
(B) Avoidance achieved by left swerving.
(C) and (D) Conflict.
}}

Now we introduce evolutionary game dynamics to the model.
Each particle has preference of right-swerving and left-swerving, which are described with $P^{\rm{R}}_i \in \bbR_{\ge 0}$ and $P^{\rm{L}}_i \in \bbR_{\ge 0}$, respectively.
Note that the superscripts R and L represent right swerving and left swerving, respectively, in the following.
The right-swerving probability, which is introduced in the previous paragraph is represented by the Logit model \cite{Hausman1984} with theses preferences:
\EQL{eq:pAvoidR}{
p_i(t) = \frac{\exp \SL P^{\rm{R}}_i(t) \SR}{\exp \SL P^{\rm{R}}_i(t) \SR + \exp \SL P^{\rm{L}}_i(t) \SR}.
}
The preferences are updated every time steps by the following equation
\footnote{
Upper bound on the preferences $P^{\rm{R}}_i(t)$ and $P^{\rm{L}}_i(t)$ become $1 / \phi$ in our model.
}
:
\EQL{eq:preference}{
P^X_i(t+1) = (1-\phi)P^X_i(t) + S^X_i(t),
}
where $X \in \{ \rm{R}, \rm{L} \}$, $\phi \in (0, 1]$ is the memory-loss rate, and $S^X_i(t)$ is the payoff for the particle $i$ at the time step $t$. 
The payoff $S^{\rm{R}}=1$ when the particles succeed in avoiding conflict by swerving to the right (Case (A) in Figs. \ref{f5_swerve2} and \ref{f4_swerve1}).
Similarly, $S^{\rm{L}}=1$ when the particles succeed in avoiding conflict by swerving to the left (Case (B) in Figs. \ref{f5_swerve2} and \ref{f4_swerve1}).
In the other cases, $S^{\rm{R}}=S^{\rm{L}}=0$.

Therefore, if the particles often interact with the opponent particles and succeed in avoiding, their preferences increase.
By contrast, if they fail to avoid the opponent particles, their preferences do not increase.
Furthermore, when there are few interaction, the preferences decrease due to the memory-loss rate $\phi$.

\SEC{Symmetric case}{EQUAL}
Here, we consider the case where the same number of right-going and left-going particles are moving in the system, i.e., $\rho_{\rm{R}}=\rho_{\rm{L}}(\equiv \rho)$.
We control the density of the particles $\rho$ and the memory-loss rate $\phi$, and investigate the two quantities.

The first one is the unified ratio defined as follows:
\EQ{
U = \left| \frac{ \sum_{i=1}^{N} 2 (p_i - 1/2) }{N} \right| \in [0,1].
}
$U \approx 1$ implies that the unified phase is achieved, i.e., most of the particles swerve to the same direction when they face their opponent particles.
On the other hand, $U \approx 0$ indicates that the disordered phase is attained, i.e., most of the particles do not have their preferred swerving direction, in other words, they swerve to the right and left with the equal probability $1/2$.

The other is the flow of the particles.
The flow of right-going (left-going) particles is the average number of right-going (left-going) particles that move in one time step divided by $L$.
We describe the flows of right-going particles, left-going particles, and their sum as $J_{\rm{R}}$, $J_{\rm{L}} \in [0, 0.5]$ and $J \in [0, 1]$, respectively. 

We set the length of the system as $L=50$ and the initial preferences $P^R_i(0)=100$, $P^L_i(0)=0$.
Simulation has been conducted for 110000 time steps, and the results from $t=10001$ to $110000$ are used to calculate the average unified ratio and flow.
Note that the stationary state is achieved at $t=10001$.

\FIGT{\EPSW{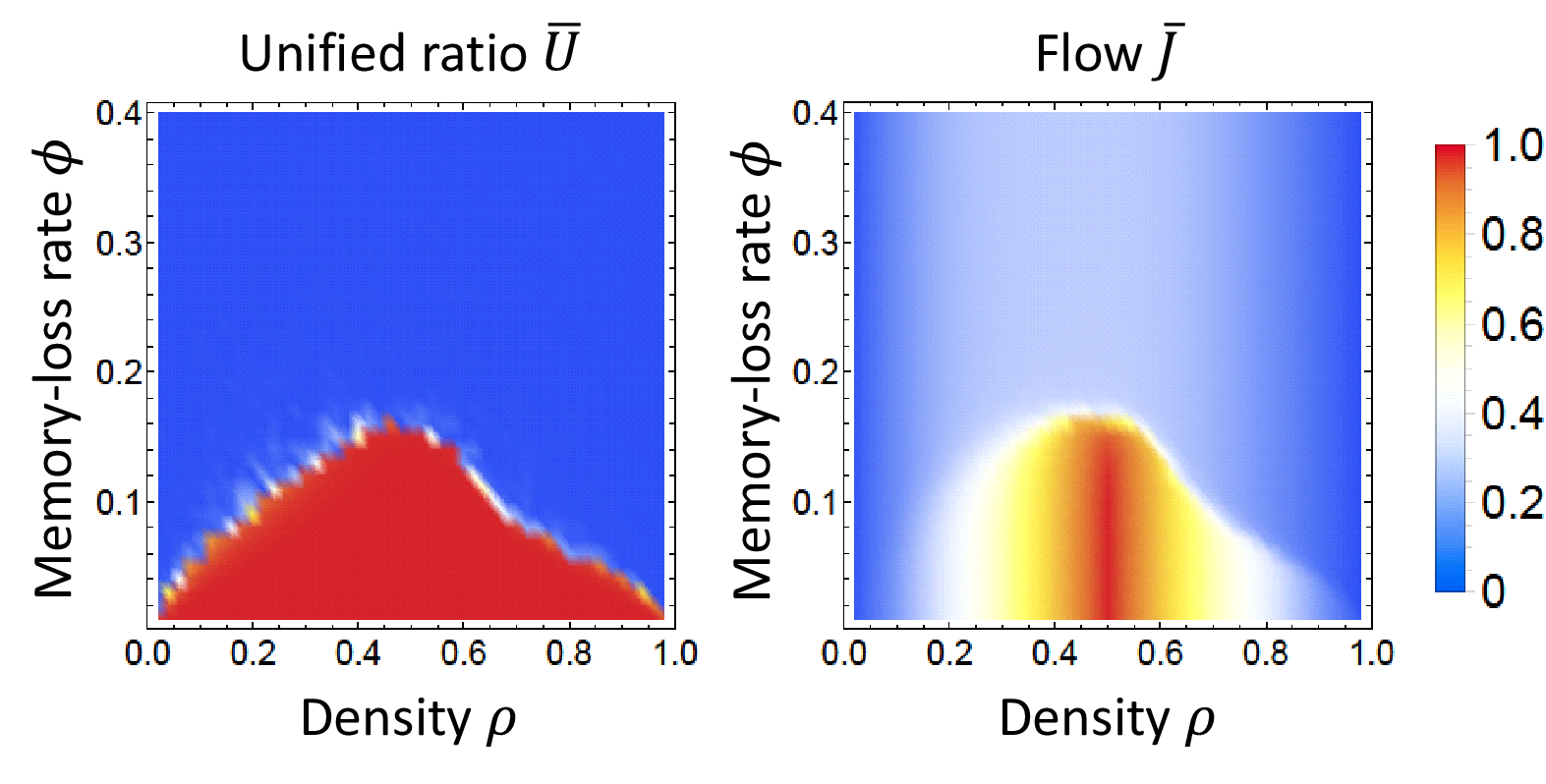}{14}{
(Left)
Average unified ratio $\bar{U}$ as a function of the density $\rho(=\rho_{\rm{R}}=\rho_{\rm{L}})$ and memory-loss rate $\phi$.
We see two clear phases, which are the disordered phase (upper blue region) and the unified phase (lower red region). 
(Right)
Average total flow $\bar{J}$ as a function of the density $\rho(=\rho_{\rm{R}}=\rho_{\rm{L}})$ and memory-loss rate $\phi$.
We see that high flow is achieved in the unified phase in the left figure.
The parameters are set as $L=50$, $P^{\rm{R}}_i(0)=100$, $P^{\rm{L}}_i(0)=0$, and the data from $t=10001$ to $110000$ are used to depict the figures.
}}

Figure \ref{NetLogo_Eq_U-Flow} (left) shows the average unified ratio $\bar{U}$ as a function of the density $\rho$ and memory-loss rate $\phi$.
We see two phases, which are the disordered (upper blue region) and unified (lower red region) phases, and phase transition between them.
When $\phi$ is large, quick memory-loss prevents the particles from keeping their preferences large, so that the disordered phase is achieved.
Even if the memory-loss rate is small, the disordered phase is observed in the low and high density region.
This is because there are few interactions between particles, which are opportunities to increase the preferences, in the low and high density cases.
In the low density case, there are few particles to interact.
In the high density case, it is difficult to move and interact since the cells are occupied by the other particles moving in the same direction. 
If the memory-loss rate is small and the density is medium, the unified phase is achieved.
Many interactions between the particles reinforce their preference. 

Figure \ref{NetLogo_Eq_U-Flow} (right) shows the average total flow $\bar{J}$ as a function of the density $\rho$ and memory-loss rate $\phi$.
We see that $\bar{J}$ achieves high values in the unified phase, while it becomes small in the disordered phase.

Cross-section diagrams of Fig. \ref{NetLogo_Eq_U-Flow} (right) at $\phi = 0.06$ and 0.30 are depicted in Fig. \ref{NetLogo_Eq_FlowTheo}.
From Fig. \ref{NetLogo_Eq_U-Flow} (left), we find that the unified phase is achieved in the most density region for $\phi=0.06$, whereas the disordered phase is attained in all the density region for $\phi=0.30$.
Together with the result of the simulation, the curves, which correspond to the double of the flow of the totally asymmetric simple exclusion process (TASEP) with the parallel update rule \cite{Schadschneider1993}, are shown in the figure.
The explicit formulation is described as
\EQ{
J = 2 \times \frac{1-\sqrt{1-4q\rho(1-\rho)}}{2},
}
where $q$ is the hopping probability of the particles and the number $2$ is multiplied because $J$ is the sum of $J_{\rm{R}}$ and $J_{\rm{L}}$.
The higher and lower curves are the flow of the TASEP in the case $q=1.0$ and 0.5, respectively.

The flows in the unified phase ($\phi=0.06$, $\rho \le 0.80$) are close to the higher curve and those in the disordered phase ($\phi=0.06$, $\rho \ge 0.86$ and $\phi=0.30$) are close to the lower curve.
Thus, the probability of successful avoidance in the unified phase and disordered phase ($\rho \ge 0.5$) in our model approximately corresponds to the hopping probability $q=1.0$ and 0.5 in the TASEP, respectively.
In the disordered phase ($\rho \le 0.5$) both the movement with the probability 1 (to the vacant cell) and 1/2 (interaction with the opponent particles) are included, so that the flow is not simply represented by the TASEP.

\FIGT{\EPSH{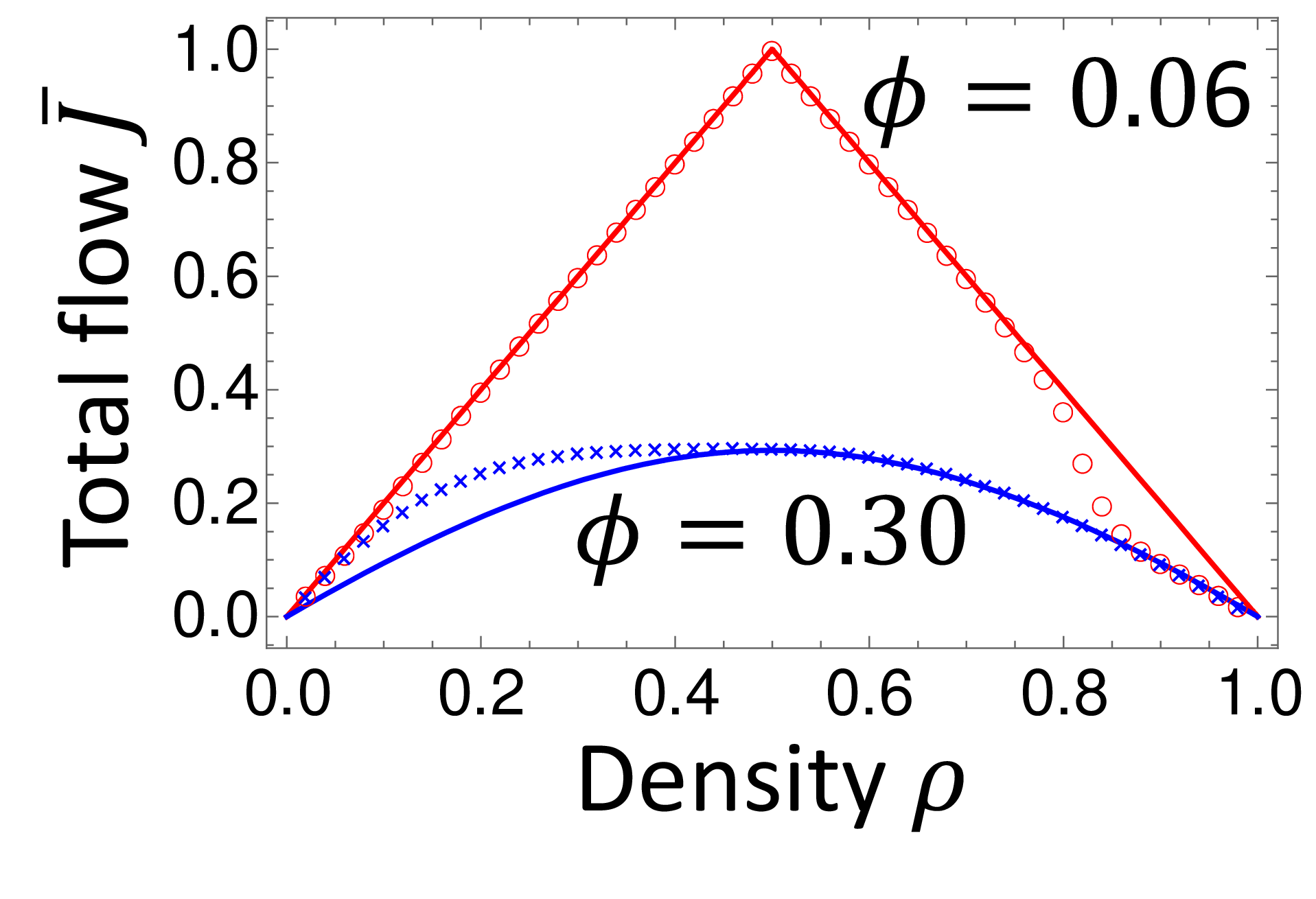}{7}{
Cross-section diagrams of Fig. \ref{NetLogo_Eq_U-Flow} (right) at $\phi = 0.06$ and 0.30.
We see that the flow of the simulation ($\phi=0.06$) agrees well with that of the TASEP ($q=1.0$) in the most part of the curve ($\rho \le 0.8$). 
By contrast, the flow of the simulation ($\phi=0.30$) agrees well with that of the TASEP ($q=0.5$) in the high density region ($\rho \ge 0.5$).
The parameters are set as $L=50$, $P^{\rm{R}}_i(0)=100$, $P^{\rm{L}}_i(0)=0$, and the data from $t=10001$ to $110000$ are used to depict the figures.
}}

\SEC{Asymmetric case}{DIFF}
Next, we consider asymmetric cases, where the number of right-going and left-going particles are different.
Fig. \ref{NetLogo_Df_U-Flow} (left) shows the average unified ratio $\bar{U}$ as a function of the density of right-going particles $\rho_{\rm{R}}$ and left-going particles $\rho_{\rm{L}}$. 
The unified phase is formed at the center of the figure.
When $\rho_{\rm{R}} \approx \rho_{\rm{L}} \approx 1/2$, both particles can move and have enough chances to interact with opponent particles.
By contrast, when $\rho_{\rm{R}}$ and $\rho_{\rm{L}}$ are greatly different, minor particles have many chances to interact, while major particles have few chances.
Thus, the preferences of major particles decrease due to the memory-loss effect, and the unified phase is collapsed.

Figure \ref{NetLogo_Df_U-Flow} (right) shows the average total flow $\bar{J}$ as a function of the density of right going particles $\rho_{\rm{R}}$ and left-going particles $\rho_{\rm{L}}$.
Similar to the symmetric case, $\bar{J}$ achieves high and low values in the unified and disordered phases, respectively.

\FIGT{\EPSW{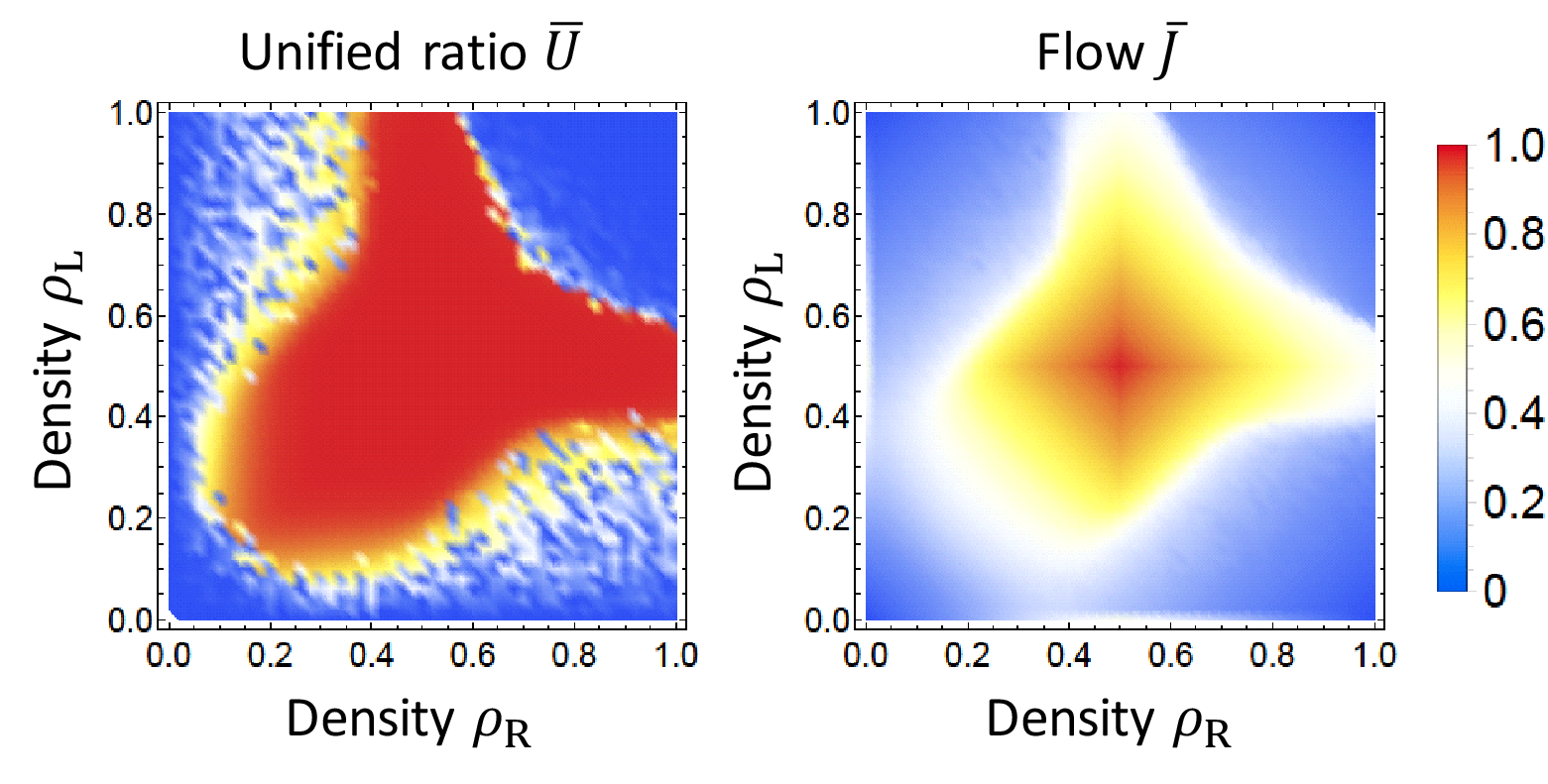}{14}{
(Left)
Average unified ratio $\bar{U}$ as a function of the density of right going particles $\rho_{\rm{R}}$ and left-going particles $\rho_{\rm{L}}$. 
We observe the unified phase (red region) at the center and the disordered phase (blue region) in the marginal part.
(Right)
Average total flow $\bar{J}$ as a function of the density of right going particles $\rho_{\rm{R}}$ and left-going particles $\rho_{\rm{L}}$.
Similar to the symmetric case, $\bar{J}$ achieve high values in the unified phase.
The parameters are set as $L=50$, $\phi=0.08$, and $P^{\rm{R}}_i(0)=100$, $P^{\rm{L}}_i(0)=0$, and the data from $t=10001$ to $110000$ are used to depict the figures.
}}

Figure \ref{NetLogo_Df_FlowCS} shows average flow of right-going particles $\bar{J_{\rm{R}}}$ as a function of $\rho_{\rm{L}}$ for various $\rho_{\rm{R}}$.
The flow $\bar{J_{\rm{R}}}$ changes non-monotonically against the increase of $\rho_{\rm{L}}$.
Firstly, in the case of $\rho_{\rm{R}}=0.2$, $\bar{J_{\rm{R}}} \approx 0.2$ for $\rho_{\rm{L}} < 0.5$, then $\bar{J_{\rm{R}}}$ drops due to the large number of opponent particles for $\rho_{\rm{L}} > 0.5$.
Secondly, in the case of $\rho_{\rm{R}}=0.5$, $\bar{J_{\rm{R}}} = 0.5$ at $\rho_{\rm{L}} = 0.0$ because there is no obstruction by left-going (opponent) particles at all.
Then it suddenly drops by one left-going particle at $\rho_{\rm{L}} = 0.02$.
For right-going particles there are few chances to interact with left-going particles, so that the disorder phase is achieved.
However, more increase of left-going particles recovers $\bar{J_{\rm{R}}}$.
Left-going particles do not only obstruct the movement of right-going particles but also increase chances of interaction for right-going particles.
Therefore, the unified phase is achieved for $\rho_{\rm{L}} > 0.3$ and $\bar{J_{\rm{R}}}$ becomes larger.
Finally, we would like to investigate the case of $\rho_{\rm{R}}=0.8$.
Similar to the case of $\rho_{\rm{R}}=0.5$, $\bar{J_{\rm{R}}}$ achieves the maximum at $\rho_{\rm{L}} = 0.0$, drops at $\rho_{\rm{R}} = 0.02$, and recovers around $\rho_{\rm{L}} \approx 0.4$.
The different phenomenon is observed around $\rho_{\rm{L}} \approx 0.7$.
In the case of $\rho_{\rm{R}}=0.8$, right-going particles cannot move smoothly, so that left-going particles have to move and increase the number of interaction.
Further increase of left-going particles deprives the mobility from them.
As a result, neither right-gong nor left-going particles smoothly move in order to interact with their opponent particles.
Hence, the unified phase is collapsed and the disordered phase is formed with the second drop of $\bar{J_{\rm{R}}}$.

\FIGT{\EPSH{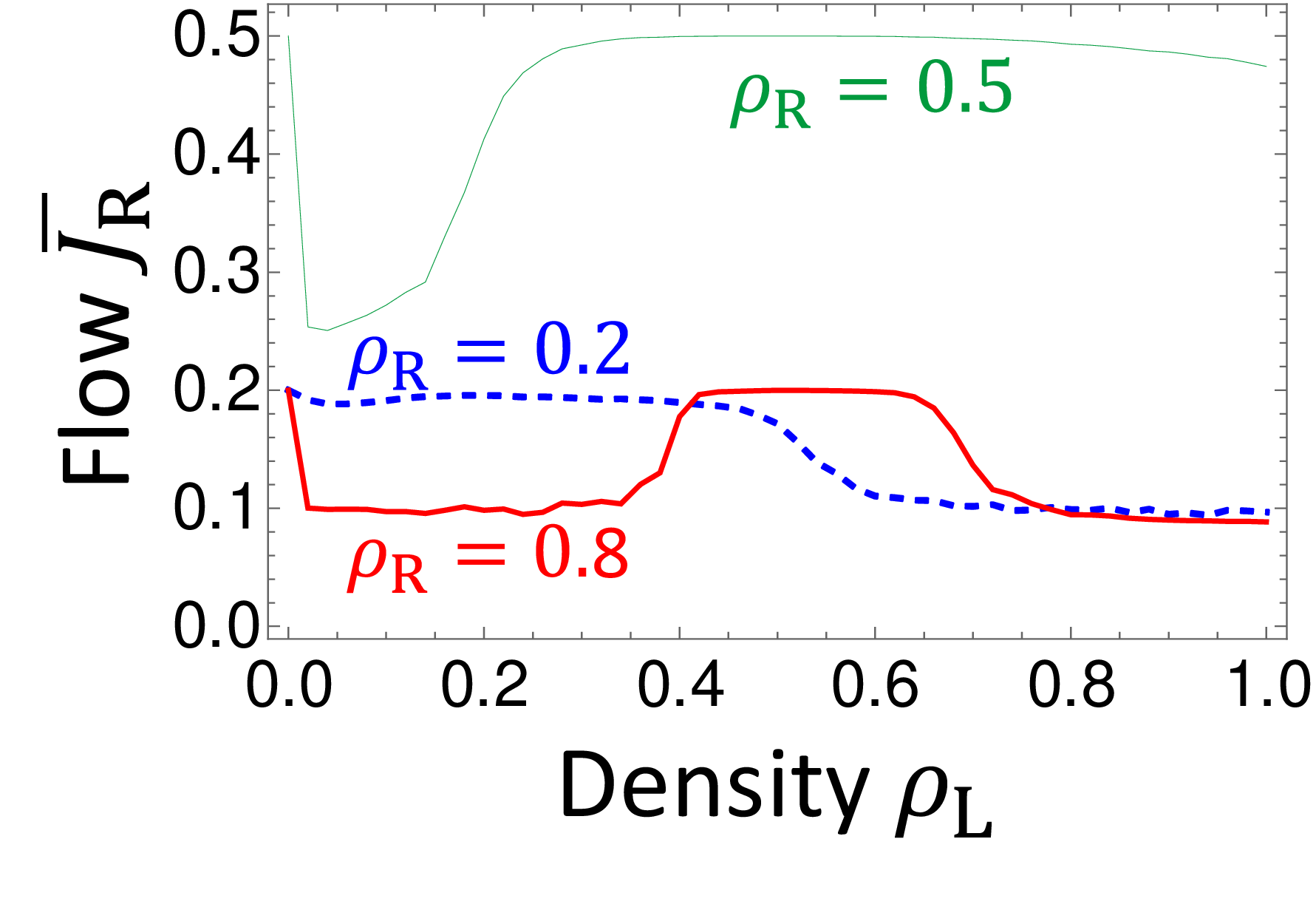}{7}{
Average flow of right-going particles $\bar{J_{\rm{R}}}$ as a function of $\rho_{\rm{L}}$ for $\rho_{\rm{R}}=0.2$, 0.5 and 0.8.
The parameters are set as $L=50$, $\phi=0.08$, and $P^{\rm{R}}_i(0)=100$, $P^{\rm{L}}_i(0)=0$, and the data from $t=10001$ to $110000$ are used to depict the figures.
}}

\SEC{Effect of system size}{SIZE}

In this section, we investigate the effect of the system size $L$ on the unified ratio $U$ and preferences $P^{\rm{R}}$ and $P^{\rm{L}}$.
Before showing the result of the simulation, we derive an approximate theoretical result for comparison.
By assuming that the properties of all the particles are always identical ($P^{\rm{R}}_i = \hat{P}^{\rm{R}}$, $P^{\rm{L}}_i = \hat{P}^{\rm{L}}$, and $p_i = \hat{p}$ for all $i$) and all the particles interact at every time step, we deform (\ref{eq:pAvoidR}) and (\ref{eq:preference}) as follows:
\EQN{
\hat{p}(t) &=& \frac{\exp \SL \hat{P}^{\rm{R}}(t) \SR}{\exp \SL \hat{P}^{\rm{R}}(t) \SR + \exp \SL \hat{P}^{\rm{L}}(t) \SR}, \label{eq:pAvoidR-Homo} \\
\hat{P}^{\rm{R}}(t+1) &=& (1-\phi)\hat{P}^{\rm{R}}(t) + (\hat{p}(t))^2, \label{eq:preferenceR-Homo} \\
\hat{P}^{\rm{L}}(t+1) &=& (1-\phi)\hat{P}^{\rm{L}}(t) + (1 - \hat{p}(t))^2, \label{eq:preferenceL-Homo} .
}
In the stationary state ($t \rightarrow \infty$), these equations are numerically solved.

Figure \ref{NetLogo_L_U} (left) shows the average unified ratio $\bar{U}$ as a function of the memory-loss rate $\phi$ for various system sizes $L=2$, 6, 50 and 1000 obtained from our simulation.
Initial preferences are set as $P^R_i(0) = 100$ and $P^L_i(0) = 0$, thus, the unified phase with right swerving is tend to be achieved.
The black curves represent the approximate results computed with (\ref{eq:pAvoidR-Homo}) - (\ref{eq:preferenceL-Homo}).
We observe the drop of $\bar{U}$ from $\bar{U}=1$ (unified phase) to $\bar{U}=0$ (disordered phase) at the critical memory-loss rate for each $L$.
The critical memory-loss rate becomes smaller as $L$ increases.
Since the dynamics of the model is stochastic, deviation from the right swerving more likely to occur when the number of particles is large.
Left swerving of one particle hinders the movement of all the particles in the system and decreases the chances of interaction.
As a result, the system becomes the disordered phase.
Therefore, it is difficult to maintain the unified phase in the large system.

Figure \ref{NetLogo_L_U} (right) shows the standard deviation of the right swerving probability $p_i$.
By comparing the left and right figures, we see that the rise of the standard deviation corresponds to the drop of $\bar{U}$.
Since the approximate theoretical result assume homogeneous properties of the particles, large standard deviation of $p_i$ deteriorate this assumption.
Thus, we see great discrepancy around $\rho=0.2$ to 0.5 in the left figure. 

\FIGT{\EPSW{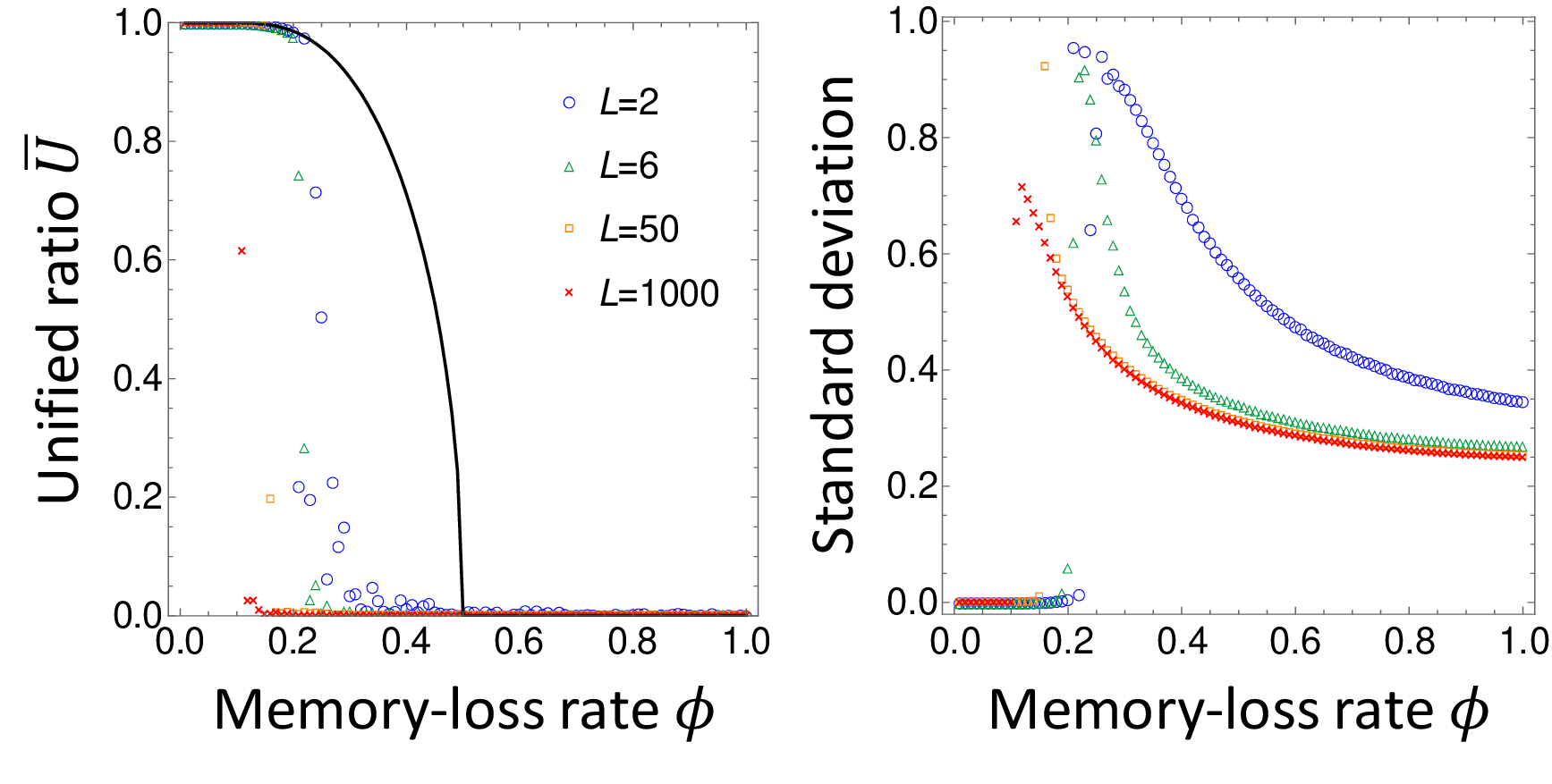}{14}{
(Left)
Average unified ratio $\bar{U}$ as functions of the memory-loss rate $\phi$ for various system sizes.
(Right)
Standard deviation of the right swerving probability $p_i$ as functions of the memory-loss rate $\phi$ for various system sizes.
The parameters are set as $\rho_{\rm{R}} = \rho_{\rm{L}} = 0.5$, $P^{\rm{R}}_i(0)=100$, and $P^{\rm{L}}_i(0)=0$.
Data from $t=10001$ to $110000$ are used for $L=2$, 6, and 50, and those from $t=2001$ to $22000$ are used for $L=1000$.
}}

In order to elucidate the large standard deviation of $p_i$, we also investigate the average preferences $\LRA{\overline{P^{\rm{R}}}}$ and $\LRA{\overline{P^{\rm{L}}}}$ as functions of the memory-loss rate $\phi$ for various system sizes.
In Fig. \ref{NetLogo_L_PRL}, $\LRA{\overline{P^{\rm{R}}}}$ and $\LRA{\overline{P^{\rm{L}}}}$ for $L=2$, 6, 50, and 1000 obtained from our simulation are depicted together with the black curves, which represent the approximate results computed with (\ref{eq:pAvoidR-Homo}) - (\ref{eq:preferenceL-Homo}).
There is only one black curve for $\phi > 0.5$, while there are two black curves for $\phi < 0.5$, which represent the larger and smaller preferences, respectively.
Since we set $P^{\rm{R}}_i(0)=100$, and $P^{\rm{L}}_i(0)=0$, the larger and smaller curves correspond to $\LRA{\overline{P^{\rm{R}}}}$ and $\LRA{\overline{P^{\rm{L}}}}$, respectively.

In Fig. \ref{NetLogo_L_PRL} (left, main), we see that all the results of simulation agree well with the approximate curve when the memory-loss rate $\phi$ is small or large.
These agreements indicate that all the particles interact every time step and maintain large $P^{\rm{R}}$ to achieve the unified phase for small $\phi$.
For large $\phi$, all the particles cannot keep $P^{\rm{R}}$, so that the disordered phase is attained.
In such region $\LRA{\overline{P^{\rm{R}}}}$ decrease with power law with the exponent -1 as the interpolated double logarithmic plot shows.
In the middle region, we observe great discrepancy between the results of the simulation and the approximate analysis similar to the unified ratio.
This phenomenon is explained with Fig. \ref{NetLogo_L_PRL} (right).
We see the rise of $\LRA{\overline{P^{\rm{L}}}}$ at the same point as the drop of $\LRA{\overline{P^{\rm{R}}}}$.
As the memory-loss rate $\phi$ increases, cooperative avoidance with left serving is sometimes succeeded in spite of the initial condition $P^{\rm{R}}_i(0)=100$ and $P^{\rm{L}}_i(0)=0$.
Success of the cooperative avoidance with left serving deteriorate the unified phase achieved with the right swerving.
In the approximate analysis, no stochastic effect is introduced, so that both $\LRA{\overline{P^{\rm{R}}}}$ and $\LRA{\overline{P^{\rm{L}}}}$ gradually changes according to $\phi$.
However, in the model, a little stochastic disturbance is enough to collapse the unified phase and form the disordered phase.
Therefore, $\bar{U}$ and $\LRA{\overline{P^{\rm{R}}}}$ in the simulation drop at much smaller $\phi$ than those of  the approximate analysis.

\FIGT{\EPSW{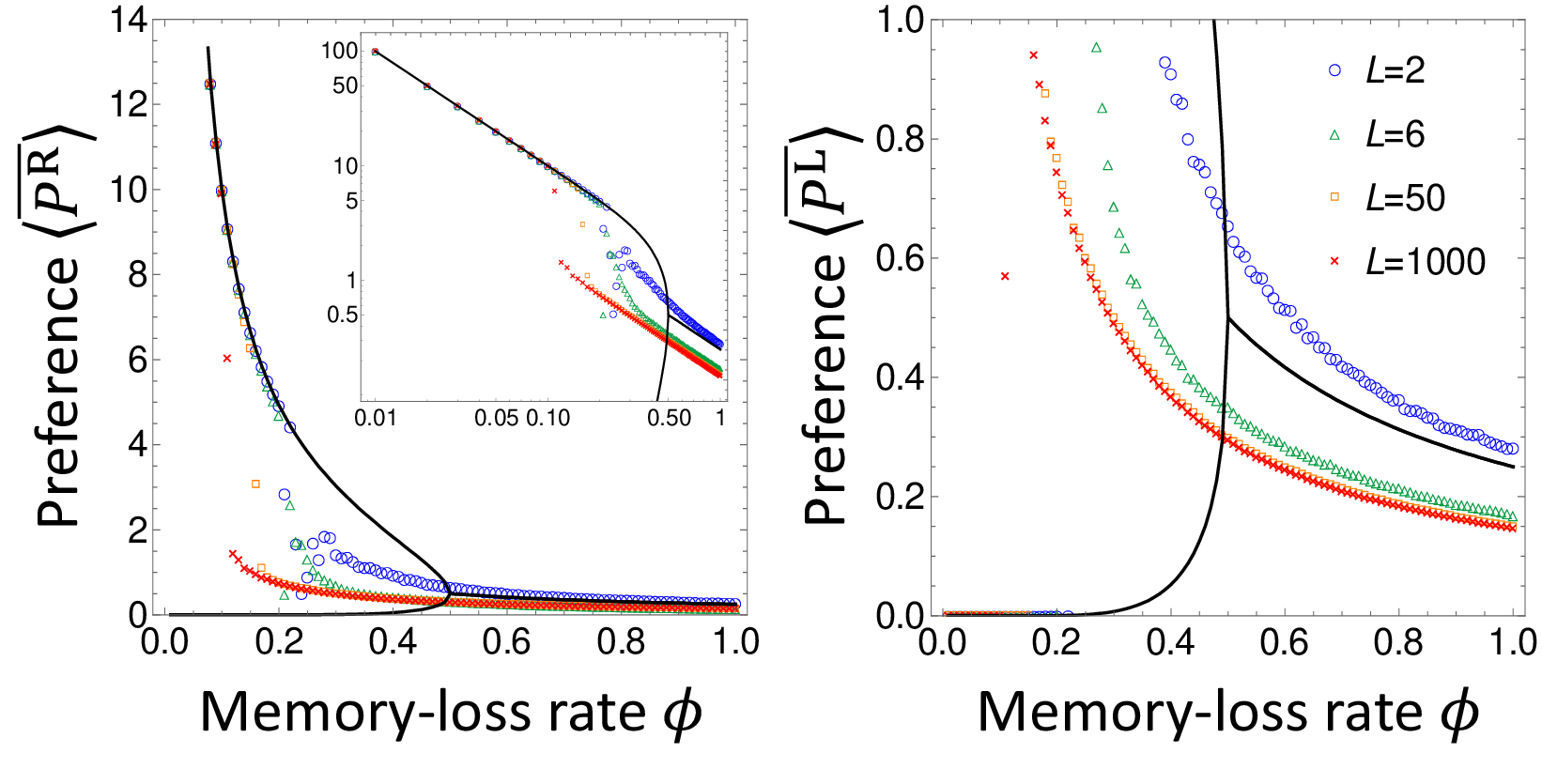}{14}{
(Left)
Average preference for right swerving $\LRA{\overline{P^{\rm{R}}}}$ as functions of the memory-loss rate $\phi$ for various system sizes.
(Right)
Average preference for left swerving $\LRA{\overline{P^{\rm{L}}}}$ as functions of the memory-loss rate $\phi$ for various system sizes.
The parameters are set as $\rho_{\rm{R}} = \rho_{\rm{L}} = 0.5$, $P^{\rm{R}}_i(0)=100$, and $P^{\rm{L}}_i(0)=0$.
Data from $t=10001$ to $110000$ are used for $L=2$, 6, and 50, and those from $t=2001$ to $22000$ are used for $L=1000$.
}}

\SEC{Summary and Conclusion}{CONC}
In this paper, we have developed a one-dimensional cellular automaton model with two kinds of particles, which are right and left-going ones.
They try to avoid each other by swerving to the right or left stochastically.
Evolutionary game dynamics is introduced in the model, so that the particles update their preferences of swerving direction by interacting other particles.
The effect of memory-loss is also considered.

The result of our simulation indicates that the swerving directions of the particles are unified (the unified phase is achieved) when the effect of memory-loss is weak and there are enough interactions between particles to increase the preference of swerving direction.
If the condition in the previous sentence is not satisfied, the serving directions are not unified (the disordered phase is achieved), in other words, all the particles swerve to the right and left with the equal probability 1/2.
It is also elucidated that the flow is well approximated by the totally asymmetric simple exclusion process (TASEP) with periodic boundary condition and the parallel-update rule.
The flows in the unified phase and disordered phase in the high density region correspond to those in the TASEP with the hopping probability equals to 1 and 1/2, respectively.
Furthermore, we investigate that the opponent particles work as both obstruction and lubricant.
When the density of the opponent particles is not adequate to the density of the main particles, they just obstruct the main flow.
However, if the density of the opponent particles is adequate, they enhance unification of the swerving direction and achieve high main flow.
The effect of the size of the system has been also studied.
We have clarified that it is difficult to maintain the unified phase in the large system.

Although the model is too simple to directly apply to the real pedestrian flow, we hope that indication from our investigation helps us to understand the mechanism of cooperating avoiding behavior in bidirectional flow.


\appendix

\SEC{Learning from failure}{LFF}

\TABLEB{
\NAME{t_lff}{
Value of the payoff for the particle $i$ $\SL S^{X}_i \SR$ as a function of the swerving directions of the particle $i$ and $j$.
The letters R and L represent right swerving and left swerving, respectively.
}
\begin{tabular}{cccc}
\hline \hline
Swerving direction of $i$ & Swerving direction of $j$ & $S^{\rm{R}}_i$ & $S^{\rm{L}}_i$ \\
\hline \hline
R & R & 1 & 0 \\
\hline
\multirow{2}{*}{R} & \multirow{2}{*}{L} & \multirow{2}{*}{0} & 1 with $p_{\rm{lff}}$ \\
& & & 0 with $1 - p_{\rm{lff}}$ \\
\hline
\multirow{2}{*}{L} & \multirow{2}{*}{R} & 1 with $p_{\rm{lff}}$ & \multirow{2}{*}{0} \\
& & 0 with $1 - p_{\rm{lff}}$ & \\
\hline
L & L & 0 & 1 \\
\hline \hline
\end{tabular}
}

In our model introduced in Sec. \ref{MODEL}, the particles increase their preferences when they succeed in avoiding conflicts.
Here, we would like to consider the particles whose preferences increase also in the case of failure in avoiding conflicts.
In other words, the particles in this appendix learn from failure.

We generalize the payoff $S^{X}_i(t)$ as in Tab. \ref{t_lff} by introducing the probability of learning form failure $p_{\rm{lff}}$.
When the particle $i$ swerve to the right and the particle $j$ swerve to the left, they fail to avoid a conflict.
Then the particle $i$ learns from this failure and increases its preference of left-swerving $P^{\rm{L}}_i$ by 1 with the probability $p_{\rm{lff}}$.
Similarly, the particle $j$ increases its preference of right-swerving $P^{\rm{R}}_j$ by 1 with the probability $p_{\rm{lff}}$.
Note that the model becomes the original one introduced in Sec. \ref{MODEL} when we set $p_{\rm{lff}} = 0$.

Figure \ref{NetLogo_Plff_U-Flow} shows the average unified ratio $\bar{U}$ and the flow $\bar{J}$ as functions of the density $\rho(= \rho_{\rm{R}} = \rho_{\rm{L}})$ and memory-loss rate $\phi$.
Actually, we see no clear difference between Figs. \ref{NetLogo_Eq_U-Flow} $(p_{\rm{lff}}=0)$ and \ref{NetLogo_Plff_U-Flow} $(p_{\rm{lff}}=1)$, so that the effect of learning from failure does not greatly influence on the stationary state of the system.
However, the effect of learning from failure on relaxation process is remained as a future work.

\FIGT{\EPSW{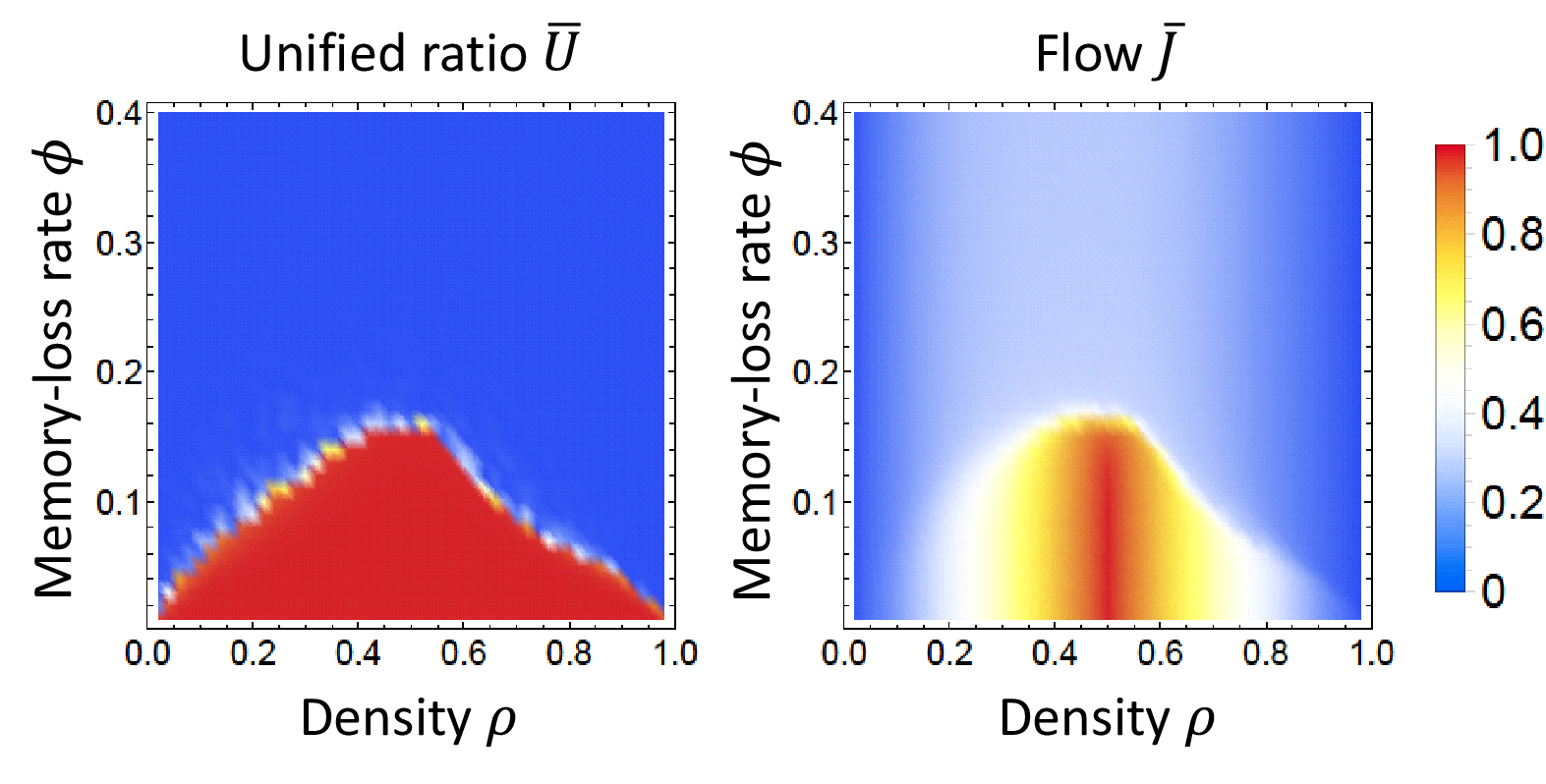}{14}{
(Left)
Average unified ratio $\bar{U}$ as a function of the density $\rho(=\rho_{\rm{R}}=\rho_{\rm{L}})$ and memory-loss rate $\phi$.
(Right)
Average total flow $\bar{J}$ as a function of the density $\rho(=\rho_{\rm{R}}=\rho_{\rm{L}})$ and memory-loss rate $\phi$.
The parameters are set as $L=50$, $P^{\rm{R}}_i(0)=100$, $P^{\rm{L}}_i(0)=0$, $p_{\rm{lff}}=1$, and the data from $t=10001$ to $110000$ are used to depict the figures.
}}

\end{document}